\documentclass[12pt]{article}

\usepackage{latexsym}

\usepackage{graphicx}

\begin{document}


\title{Dilaton black holes with squashed horizons and their thermodynamics }
\author{
     Stoytcho S. Yazadjiev \thanks{E-mail: yazad@phys.uni-sofia.bg}\\
{\footnotesize  Department of Theoretical Physics,
                Faculty of Physics, Sofia University,}\\
{\footnotesize  5 James Bourchier Boulevard, Sofia~1164, Bulgaria }\\
}

\date{}

\maketitle

\begin{abstract}
New charged solutions describing black holes with squashed horizons
in 5D dilaton gravity are presented. The black hole spacetimes are asymptotically
locally flat and have a spacial infinity $R\times S^1\hookrightarrow S^2$. The solutions
are analyzed and their thermodynamics is discussed by using the counterterm method.
\end{abstract}


\sloppy

\section{Introduction}

In recent years the higher dimensional gravity is attracting much interest. Apart from the fact
that the higher dimensional gravity is interesting on its own right, the increasing amount of works
devoted to the study of the higher dimensional spacetimes is inspired by the string theory and
the brane-world scenario with large extra dimensions \cite{AHDD}-\cite{KOKO}.
This scenario suggests a  possibility of unification of the
electro-weak and Planck scales at TeV scale. A striking prediction in this scenario is the
formation of higher dimensional black holes smaller than the size of the extra dimensions
at accelerators \cite{BF},\cite{GT}.

The growing interest in  higher dimensional gravity, and in higher dimensional black holes in particular,
reveals itself in different aspects. One of the directions of investigations is the construction
of exact  solutions both analytically and numerically \cite{T}-\cite{W}.

Some solutions of the higher dimensional classical general relativity have been known for some time. These include
the higher dimensional analogues of  Schwarzschild and Reissner-Nordstrom solution found by Tangherlini \cite{T}
and the higher dimensional generalization  of the Kerr solution found by Myers and Perry \cite{MP}.
As one should expect and as it was confirmed by recent investigations, the gravity in higher dimensions
exhibits much richer dynamics than in four dimensions.An interesting development in the black holes
studies is the discovery of the black ring solutions
of the five-dimensional Einstein equations by Emparan and Reall \cite{ER1}, \cite{ER2}. These are
asymptotically flat solutions with $S^2\times S^1$ topology of  the event horizon unlike the
much more familiar $S^3$ topology. Moreover, it was shown in \cite{ER2} that both the black hole
and the  black ring can carry the same conserved charges, the mass and a single angular
momentum, and therefore there is no uniqueness theorem in five dimensions. Since the Emparan and
Reall's discovery many explicit examples of black ring solutions were found and discussed  in various gravity
theories \cite{E}-\cite{MI2}. In \cite{EMP}
Emparan derived "dipole black rings" in Einstein-Maxwell-dilaton (EMd) theory
in five dimensions. In this work Emparan showed that the black rings can exhibit novel
features with respect to the black holes. Black rings can also carry nonconserved charges
which can be varied continuously without altering the conserved charges. This fact leads to
continuous non-uniqness. The systematic derivation of the dipole black  ring solutions was
given in the works of the author \cite{Y4} and \cite{Y5}.
The thermodynamics of the dipole black rings was studied first by Emparan in \cite{EMP} and by
Copsey  and Horowitz in \cite{CH}. Within the framework of the quasilocal
counterterm method, the thermodynamics of the dipole rings was discussed by Astefanesei and Radu \cite{AR}.
The first law of black rings thermodynamics in $n$-dimensional Einstein dilaton gravity with $(p+1)$-form field
strength was derived by Rogatko in \cite{ROG}(see also \cite{ROG1} and  \cite{ROG2}).

Recently, Ishihara and Matsuno \cite{IM} found new black hole solutions with unusual topology in 5D Einstein-Maxwell (EM) gravity--the so-called
black holes with squashed horizons. These black holes have horizons in the shape of squashed $S^3$, and their asymptotic structure
is that of the fiber bundle $R\times S^1 \hookrightarrow S^2$. The spacetime that looks like a five-dimensional black hole in the vicinity of the hole gives an effective four-dimensional black hole with a compact extra dimension at infinity. The thermodynamics of the Ishihara-Matsuno black holes was discussed in \cite{CCO}. Rotating black hole solutions with squashed horizons in 5D Einstein gravity were found and discussed by Wang in \cite{W}.

In the present paper we present new charged black hole solutions with squashed horizons in 5D Einstein-Maxwell-dilaton (EMd) gravity.
The solutions are analyzed and their thermodynamics is discussed.

\section{A review of Einstein-Maxwell black holes  with  \\ squashed horizons}

In this section we give a brief review of the black hole solutions of the 5D Einstein-Maxwell theory
with squashed horizons. The metric is given by

\begin{eqnarray}
ds^2 = - fdt^2 + {k^2 \over f} dr^2 + {r^2\over 4 }  \left[k\left((\sigma^{1})^2 + (\sigma^{2})^2 \right) + (\sigma^{3})^2\right]
\end{eqnarray}

where $f$ and $k$ are functions of r defined by

\begin{eqnarray}
f(r) &=& {(r^2- r^2_{+})(r^2- r^2_{-}) \over r^4 } ,\\
k(r) &=&  {(r^2_{\infty}- r^2_{+})(r^2_{\infty}- r^2_{-}) \over (r^2_{\infty}- r^2)^2 } .
\end{eqnarray}

The gauge potential is

\begin{equation}
A_{t} = \pm {\sqrt{3}\over 2} {r_{+}r_{-}\over r^2 } + const.
\end{equation}

Here, $r_{\pm}$ and $r_{\infty}$ are constants with $0<r_{-}\le r_{+} <r_{\infty}$.
The radial coordinate $r$ takes values in the range

\begin{equation}
0< r< r_{\infty}.
\end{equation}

Respectively,

\begin{eqnarray}
\sigma^{1} &=& - \sin\psi d\theta + \cos\psi\sin\theta d\phi ,\\
\sigma^{2} &=& \cos\psi d\theta + \sin\psi\sin\theta d\phi ,\\
\sigma^{3} &=& d\psi + \cos\theta d\phi ,
\end{eqnarray}

are Maurer-Cartan one form on $S^{3}$ and $\theta$, $\phi$ and $\psi$ are Euler angles with

\begin{equation}
0<\theta <\pi, \,\,\, 0<\phi <2\pi, \,\,\, 0<\psi < 4\pi .
\end{equation}

The spacetime metric has the isometry group $SO(3)\times U(1)$. It also has an outer horizon at $r=r_{+}$ and
 inner one at $r=r_{-}$. There is an inner singularity at $r=0$. Time-slices $t=const$, which are orthogonal to the Killing vector
$\partial/\partial t$, are foliated by the three-dimensional surfaces $\Sigma_{r}$ defined by $r=const$. Each surface $\Sigma_{r}$ has the topology of the Hopf fiber bundle $S^1 \hookrightarrow S^2$
with the metric

\begin{eqnarray}
ds^2_{\Sigma_{r}} = {r^2\over 4} \left[k(r)\left((\sigma^{1})^2 + (\sigma^{2})^2 \right) + (\sigma^{3})^2 \right] =
{r^2\over 4} \left[k(r)d\Omega^2_{S^2} + \chi^2  \right]
\end{eqnarray}

where

\begin{eqnarray}
d\Omega^2_{S^2} &=& d\theta^2 + \sin^2\theta d\phi^2 ,\\
\chi &=&\sigma^{3} = d\psi + \cos\theta d\phi.
\end{eqnarray}

The function $k(r)$ causes the deformation of the black hole horizon. For $k(r)=1$ (i.e. $r_{\infty}\to \infty$) the
solution is just the five dimensional Reissner-Nordtsrom black hole whose  horizon is a round three-sphere $S^{3}$.

In the limit $r_{-}\to 0$  we obtain a neutral black hole which is described by the metric

\begin{equation}\label{NM}
ds^2 = - (1 -{r^2_{+}\over r^2})dt^2  + {k\over 1 - {r^2_{+}\over r^2 } } dr^2 + {r^2\over 4 }\left[kd\Omega^2_{S^2} + \chi^2\right].
\end{equation}

For finite $r_{\infty}$ the black hole metric is deformed by the squashing function $k(r)$ with $r_{-}=0$. In the limit $r_{\infty}\to \infty$
the metric (\ref{NM}) reduces to the 5D Schwarzschild black hole  with $SO(4)$ symmetry, while for $r_{+}\to 0$ the metric reduces
to Gross-Perry-Sorkin monopole \cite{GP},\cite{S}.

The spacial infinity is reached for $r=r_{\infty}$. In turns out more convenient to study the asymptotic behaviour  in
terms of the new radial coordinate $\rho$ defined by

\begin{equation}
\rho = \rho_{0} {r^2 \over r^2_{\infty}- r^2  }
\end{equation}

where

\begin{eqnarray}
\rho^2_{0} &=& k_{0} {r^2_{\infty}\over 4 } ,\\
k_{0} &=& k(r=0)= f_{\infty}=f(r=r_{\infty})= {(r^2_{\infty} - r^2_{+}) (r^2_{\infty} - r^2_{-}) \over r^4_{\infty}}.
\end{eqnarray}

The new radial coordinate varies in the range $0<\rho<\infty$. In terms of $\rho$ and the new time coordinate $\tau=\sqrt{f_{\infty}}\, t$ the
metric can be rewritten as

\begin{eqnarray}\label{RHOM}
ds^2 = - Vd\tau^2 + {K^2\over V}d\rho^2 + R^{2} d\Omega^2_{S^2} + W^2\chi^2
\end{eqnarray}

where $V$, $K$, $R$ and $W$ are functions of $\rho$ given by

\begin{eqnarray}
V &=& {(\rho- \rho_{+})(\rho-\rho_{-})\over \rho^2 }  ,\\
K^2 &=& {\rho + \rho_{0} \over \rho} ,\\
R^2 &=& \rho^2 K^2 ,\\
W^2 &=&  {r^2_{\infty}\over 4} K^{-2} = (\rho_{0} + \rho_{+})(\rho_{0} + \rho_{-}) K^{-2}.
\end{eqnarray}

The parameters $\rho_{\pm}$ are defined by

\begin{equation}
\rho_{\pm} =\rho_{0} { r^2_{\pm}\over r^2_{\infty} - r^2_{\pm} }.
\end{equation}

In the limit $\rho\to \infty$ (i.e. $r\to r_{\infty}$) the metric (\ref{RHOM}) approaches

\begin{equation}
ds^2 = - d\tau^2 + d\rho^2 + \rho^2 d\Omega^2_{S^2} + {r^2_{\infty}\over 4}\chi^2.
\end{equation}

The boundary topology is that of the fibre bundle $R\times S^1\hookrightarrow S^2$ as the radius
of $S^2$ is growing with $\rho$, while the radius of $S^1$ reaches a constant value.

\section{Dilaton black holes with squashed horizons}

In this paper we consider 5D EMd theory  with action

\begin{eqnarray}
S= {1\over 16\pi} \int d^5x \sqrt{-g}\left(R - 2g^{\mu\nu}\partial_{\mu}\varphi \partial_{\nu}\varphi  -
e^{-2\alpha\varphi}F^{\mu\nu}F_{\mu\nu} \right).
\end{eqnarray}

where $\alpha$ is the dilaton coupling parameter.  The equations of motion can be obtained by varying with respect
to the spacetime metric $g_{\mu\nu}$, dilaton field $\varphi$ and gauge potential $A_{\mu}$, which yields

 \begin{eqnarray}\label{EMDE}
R_{\mu\nu} &=& 2\partial_{\mu}\varphi \partial_{\nu}\varphi + 2e^{-2\alpha\varphi} \left[F_{\mu\rho}F_{\nu}^{\rho}
- {1\over 6} g_{\mu\nu} F_{\beta\rho} F^{\beta\rho}\right], \nonumber \\
\nabla_{\mu}\nabla^{\mu}\varphi &=& -{\alpha\over 2} e^{-2\alpha\varphi} F_{\nu\rho}F^{\nu\rho}, \\
&\nabla_{\mu}&\left[e^{-2\alpha\varphi} F^{\mu\nu} \right]  = 0 \nonumber .
\end{eqnarray}

A method for generating static solutions to the field equations (\ref{EMDE}) in $D$-dimensions was developed in \cite{Y} and \cite{Y1}.
This method  allows us to generate exact EMd solutions from known static solutions of the vacuum $D$-dimensional Einstein equations.
Our seed solution is the 5D vacuum Einstein black hole  (\ref{NM}) with redefined time coordinate so that
$g_{00}(r_{\infty})=-1$ i.e.

\begin{equation}\label{SS}
ds^2 = - {1 - {r^2_{+}\over r^2 } \over 1 - {r^2_{+}\over r^2_{\infty} }} dt^2 + {\omega^2(r) \over 1- {r^2_{+}\over r^2 } }dr^2
+ {r^2 \over 4} \left[\omega(r)d\Omega^2_{S^2}  + \chi^2 \right]
\end{equation}

where

\begin{equation}
\omega(r) = k(r; r_{-}=0) = {(r^2_{\infty} - r^2_{+} )r^2_{\infty} \over (r^2_{\infty}-r^2)^2 }.
\end{equation}

Applying the method to the seed vacuum solution (\ref{SS}) we find the following new  EMd solution

\begin{eqnarray}
ds^2 &=& - {1-{r^2_{+}\over r^2} \over 1 - {r^2_{+}\over r^2_{\infty} } }
\left[1 + {r^2_{+}\over r^2_{\infty} - r^2_{+}} {r^2_{\infty} - r^2 \over r^2}\sinh^2(\vartheta) \right]^{-{2\over 1+ \alpha^2_{*}}} dt^2
 \\ &+& \left[1 + {r^2_{+}\over r^2_{\infty} - r^2_{+}} {r^2_{\infty} - r^2 \over r^2}\sinh^2(\vartheta) \right]^{1\over 1+ \alpha^2_{*}}
 \left[ {\omega^2 \over 1 - {r^2_{+}\over r^2}  }dr^2 +  {r^2\over 4 } \left[\omega d\Omega^2_{S^2} + \chi^2 \right]  \right] ,\nonumber\\
e^{-2\alpha\varphi} &=& \left[1 +
 {r^2_{+}\over r^2_{\infty} - r^2_{+}} {r^2_{\infty} - r^2 \over r^2}\sinh^2(\vartheta) \right]^{2\alpha^2_{*}\over 1+ \alpha^2_{*}} ,\\
A_{t} &=& {\sqrt{3} \cosh(\vartheta)\sinh(\vartheta)\over 2\sqrt{1+ \alpha^2_{*}}}
{(r^2_{\infty} - r^2)r^2_{+}\over  r^2(r^2_{\infty} - r^2_{+} ) + r^2_{+} (r^2_{\infty} - r^2 )\sinh^2(\vartheta) } ,
\end{eqnarray}

where

\begin{equation}
\alpha_{*} = {\sqrt{3}\over 2} \alpha .
\end{equation}

Here $r_{+}<r_{\infty}$ and  the radial coordinate $r$ varies in the range

\begin{equation}
 0 < r< r_{\infty} .
\end{equation}

For $\alpha=0$, this solution reduces to the EM squashed black hole solution by Ishihara-Matsuno \cite{IM}.
In order to see that, we have to perform the coordinate change $r\to {\tilde r}$ where

\begin{equation}
{\tilde r}^2= r^2 + {r^2_{+} \over r^2_{\infty} - r^2_{+}} (r^2_{\infty} - r^2)    \sinh^2(\vartheta)
\end{equation}

as well as to define the new parameters

\begin{eqnarray}
{\tilde r}_{+}&=&r_{+}\cosh(\vartheta) ,\\
{\tilde r}_{-}&=&
{r_{\infty}r_{+}\over \sqrt{r^2_{\infty}-r^2_{+}} }\sinh(\vartheta) ,\\
{\tilde r}_{\infty} &=& r_{\infty}.
\end{eqnarray}.

For $r_{\infty}\to \infty$ our solution reduces to the usual 5D dilaton black hole solution with $SO(4)$ symmetry.
Let us also note that the constant in the gauge potential $A_{t}$ is fixed so that

\begin{equation}
A_{t}(r_{\infty}) = 0.
\end{equation}

The metric has an event horizon at $r=r_{+}$. The time-slice of the horizon is described by the three-metric

\begin{eqnarray}
dl^2 = [\cosh^2(\vartheta)]^{1\over 1 + \alpha^2_{*}} {r^2_{+}\over 4} \left[\omega(r_{+})d\Omega^2_{S^2} + \chi^2 \right]
\end{eqnarray}

where the horizon squashing function is

\begin{equation}
\omega(r_{+}) = {r^2_{\infty}\over r^2_{\infty} - r^2_{+}}.
\end{equation}

The horizon and the three-dimensional surfaces $r=const$ that foliate the time-slice $t=const$ have topology $S^1 \hookrightarrow S^2$. These surfaces are  the  squashed $S^3$ on which  $SO(3)\times U(1)$ acts as an isometry group.

There is an inner curvature singularity at $r=0$ -- in the vicinity of $r=0$ the Ricci scalar curvature behaves as

\begin{equation}
R \sim r^{-2{1 + 2\alpha^2_{*}\over 1+ \alpha^2_{*}}} .
\end{equation}

The spacial infinity  corresponds to $r=r_{\infty}$. In order to see that,
we introduce the new radial coordinate $\rho$ given by

\begin{equation}
\rho = {\sqrt{r^2_{\infty} - r^2_{+}} \over 2} {r^2 \over r^2_{\infty} - r^2 }
\end{equation}

which runs from $0$ to $\infty$. Let us also define

\begin{eqnarray}
\rho_{+} &=& \rho(r_{+}) = {r^2_{+}\over 2\sqrt{r^2_{\infty} - r^2_{+}} } ,\\
\rho_{0} &=& {1\over 2}\sqrt{r^2_{\infty} - r^2_{+}} .
\end{eqnarray}

Then, in terms of the new radial coordinate $\rho$ the solution takes the form

\begin{eqnarray}
ds^2 &=& - {\left(1 - {\rho_{+}\over \rho}\right)\over \left[1 + {\rho_{+}\over \rho}\sinh^2(\vartheta)\right]^{2\over 1+ \alpha^2_{*}}} dt^2
 \\ &+& \left[1 + {\rho_{+}\over \rho}\sinh^2(\vartheta)\right]^{1\over 1+ \alpha^2_{*}}
\left[{ \left(1 + {\rho_{0}\over \rho }\right)\over \left(1 - {\rho_{+}\over \rho }\right)} d\rho^2 + \rho (\rho + \rho_{0})d\Omega^2_{S^{2}} +
(\rho_{0} + \rho_{+}) {\rho_{0}\rho\over \rho + \rho_{0}}\chi^2 \right] ,\nonumber \\
e^{-2\alpha\varphi} &=& \left[1 + {\rho_{+}\over \rho}\sinh^2(\vartheta) \right]^{\alpha^2_{*} \over 1 + \alpha^2_{*}} , \\
A_{\tau} &=& {\sqrt{3}\cosh(\vartheta)\sinh(\vartheta)\over 2\sqrt{1 + \alpha^2_{*}}} {\rho_{+}\over \rho + \rho_{+}\sinh^2(\vartheta) }
\end{eqnarray}

In the limit $\rho \to \infty$ (i.e. $r\to r_{\infty}$) the metric approaches

\begin{equation}
ds^2 = - dt^2 + d\rho^2 + \rho^2 d\Omega^2_{S^2} + (\rho_{0} + \rho_{+}) \rho_{0} \chi^2.
\end{equation}

The spacetime is locally asymptotically flat with boundary topology $R\times S^1 \hookrightarrow S^2$.

\section{Thermodynamics of the dilaton black holes with squashed horizons}

To study the thermodynamics we have to find the conserved charges  of our system first. Throughout the
years many methods have been developed  to calculate the conserved charges of the gravitational configurations.
A related problem is the computation of the gravitational action of a noncompact spacetime. When evaluated on
non-compact solutions, the bulk Einstein term and the boundary Gibbons-Hawking term are both divergent.
The remedy is to consider these quantities relative to those associated with some background reference spacetime,
whose boundary at infinity has the same induced metric as that of the original spacetime. This substraction
procedure is  however connected to difficulties: the choice of the reference background is by no means unique
and  it is not always possible to embed a boundary with a given induced metric into the reference
background. New method, free of the mentioned difficulties, was proposed by Balasubramanian and Kraus \cite{BK}.
This method (called counterterm  method) consists in adding a (counter)term to the boundary at
infinity, which is a functional only of the curvature invariants of the induced metric on the boundary.
Unlike the substraction procedure, this method is intrinsic to the spacetime of interest and
is unambiguous once the counterterm which  cancels the divergencies is specified. In the present
work we use namely the counterterm  method in order to compute the mass of the dilaton black holes
with squashed horizons. Our solution has boundary topology  $R\times S^1\hookrightarrow S^2$ which is the same
as that of the Kaluza-Klein monopole \cite{MS}. In the case of Kaluza-Klein monopole, Mann and Stelea \cite{MS}
proposed the following simple counterterm

\begin{equation}
I_{ct} = {1\over 8\pi} \int d^4x \sqrt{-h} \sqrt{2{\cal R}}
\end{equation}

where ${\cal R}$ is the Ricci scalar scalar with respect to the boundary metric $h_{ij}$. With this counterterm,
the boundary stress-energy tensor is found to be

\begin{equation}
T_{ij} = {1\over 8\pi} \left[ K_{ij} - K h_{ij} - \Psi ({\cal R}_{ij} - {\cal R}h_{ij} ) - h_{ij} D^kD_{k}\Psi + D_{i}D_{j}\Psi   \right].
\end{equation}

Here $K$ is the trace of the extrinsic curvature $K_{ij}$ of the boundary, $\Psi= \sqrt{2\over {\cal R}}$ and $D_{k}$  is the covariant
derivative with respect to the metric $h_{ij}$. If the boundary geometry has  an isometry generated by the Killing vector $\xi$, $T_{ij}\xi^{j}$
is divergence free which gives the conserved quantity

\begin{equation}
{\cal Q} = \int_{\Sigma} d\Sigma_{i} T^{i}_{j}\xi^{j}
\end{equation}

associated with the closed surface $\Sigma$.  In the case when $\xi = \partial/\partial t $,    $\cal Q$ is the conserved mass.

For our dilaton solution  we have

\begin{equation}
8\pi T^{t}_{t} = {1\over \rho^2} \left[{1\over 2}\rho_{0} + \rho_{+} +  {3\over 2}{\sinh^2(\vartheta)\over 1 + \alpha^2_{*} } \rho_{+} \right]
+ {\cal O}({1\over \rho^3})
\end{equation}

Integrating we find for the black hole mass

\begin{equation}
M = 2\pi \sqrt{(\rho_{0} + \rho_{+} ) \rho_{0}} \left[{1\over 2}\rho_{0} + \rho_{+}
+ {3\over 2}{\sinh^2(\vartheta)\over 1+ \alpha^2_{*} } \rho_{+} \right].
\end{equation}

In terms of $r_{\infty}$ and $r_{+}$ the mass is given by

\begin{equation}
M = {\pi r_{\infty}\over 4\sqrt{ r^2_{\infty} - r^2_{+} } } \left[r^2_{\infty} + r^2_{+}  + {3\sinh^2(\vartheta)\over 1 + \alpha^2_{*} } r^2_{+} \right]
\end{equation}

The temperature can be derived by  continuing the metric to its Euclidean sector and requiring the absence of conical singularities at the horizon. This results in periodic Euclidean time with period $1/T$ where $T$ is the Hawking temperature. The explicit calculations give

\begin{equation}
T =  { \left[\cosh(\vartheta)\right]^{- {3\over 1+ \alpha^2_{*}}} \over 4\pi \sqrt{\rho_{+} (\rho_{0} + \rho_{+}) }} =
 { \left[\cosh(\vartheta)\right]^{- {3\over 1+ \alpha^2_{*}}} \over 2\pi } {\sqrt{r^2_{\infty} - r^2_{+} }\over  r_{\infty}r_{+}} .
\end{equation}

The area of the horizon can be found via straightforward calculation and the result is

\begin{equation}
{\cal A}_{h} = 16 \pi^2 \left[\cosh(\vartheta)\right]^{3\over 1+ \alpha^2_{*}}
\rho_{+}^{3\over 2} \rho^{1\over 2 }_{0} (\rho_{0} + \rho_{+}) = 2\pi^2   \left[\cosh(\vartheta)\right]^{3\over 1+ \alpha^2_{*}}
{r^3_{+} r^2_{\infty} \over r^2_{\infty} - r^2_{+}  } .
\end{equation}

It was found in \cite{CCO} by using  Wald's formula \cite{Wald} that the entropy of  EM black holes with squashed horizons  obey the area law although the horizons are deformed. In the same way, one can show that  the area law still holds in our case, i.e.

\begin{equation}
S= {1\over 4} {\cal A}_{h}.
\end{equation}

The electric charge defined by

\begin{equation}
Q = - {1\over 4\pi} \oint_{S^3} \star e^{-2\alpha\varphi} F
\end{equation}

is found to be

\begin{equation}
Q = 2\pi{\sqrt{3}\cosh(\vartheta)\sinh(\vartheta)\over \sqrt{1 + \alpha^2_{*}} } \rho_{+} \sqrt{\rho_{0}(\rho_{0} + \rho_{+}) } =
\pi {\sqrt{3}\cosh(\vartheta)\sinh(\vartheta)\over 2\sqrt{1 + \alpha^2_{*}} } {r_{\infty} r^2_{+}\over \sqrt{r^2_{\infty} - r^2_{+} } } .
\end{equation}

The gauge potential evaluated on the horizon is

\begin{equation}
\Phi_{h} = A_{t}(\rho_{+}) = {\sqrt{3}\tanh(\vartheta)\over 2\sqrt{1+ \alpha^2_{*}} }.
\end{equation}

It can be checked that the quantities $M$, $S$, $T$, $Q$ and $\Phi_{h}$ satisfy the first law of black hole thermodynamics

\begin{equation}
dM = TdS + \Phi_{h} dQ.
\end{equation}

It has to be noted that,  in the above formula we have considered $r_{+}$ and $\vartheta$ as variables and $r_{\infty}$ as  constant.

\section{Conclusion}

In the present paper we have presented new solutions to 5D EMd theory describing dilaton black holes
with squashed horizons. The solutions were analyzed and their masses were calculated by using the counterterm
method. The black hole thermodynamics was discussed, too. A natural generalization of this work is
the finding EMd solutions describing rotating dilaton black holes with squashed horizons. This problem
is currently under investigation and the results will be presented elsewhere.

\section*{Acknowledgements}
I would like to thank I. Stefanov for reading the manuscript.
This work was partially supported by the
Bulgarian National Science Fund under Grant MUF04/05 (MU 408)
and the Sofia University Research Fund under Grant No60.

\end{document}